\documentclass[12pt,preprint]{aastex}

\usepackage{epsfig}

\received{2004 December 30}
\begin{document}


\newcommand{\be}{\begin{equation}}
\newcommand{\ee}{\end{equation}}
\newcommand{\ba}{\begin{eqnarray}}
\newcommand{\ea}{\end{eqnarray}}
\newcommand{\siml}{\lower4pt \hbox{$\buildrel < \over \sim$}}
\newcommand{\simg}{\lower4pt \hbox{$\buildrel > \over \sim$}}
\newcommand{\Mesz}{M\'esz\'aros}
\newcommand{\epsd}{\epsilon_d}
\newcommand{\cm}{{\rm cm}}
\newcommand{\erg}{{\rm erg}}
\newcommand{\si}{{\rm s}^{-1}}



\title{Dissipative Photosphere Models of \\ Gamma-ray Bursts and X-ray 
Flashes}

\author{M.J.~Rees\altaffilmark{1} and  P. \Mesz\altaffilmark{2,3}}

\altaffiltext{1}{Institute of Astronomy, Univ. of Cambridge, Madingley Rd, Cambridge CB30HA, U.K}

\altaffiltext{2}{Dpt.Astron.\&Astrophys., Dpt. Physics,
Pennsylvania State U., University Park, PA 16803}

\altaffiltext{3}{Institute for Advanced Study, Princeton, NJ 08540}



\begin{abstract}

We consider dissipative effects occurring in the optically thick inner 
parts of the relativistic outflows producing gamma-ray bursts and X-ray 
flashes, emphasizing specially the Comptonization of the thermal radiation 
flux that is advected from the base of the outflow. Such dissipative 
effects --e.g. from magnetic reconnection, neutron decay or shocks -- 
would boost the energy density of the thermal radiation. The dissipation 
can lead to pair production, in which case the pairs create an effective 
photosphere further out than the usual baryonic one. In a slow dissipation
scenario, pair creation can be suppressed, and the effects are most
important when dissipation occurs below the baryonic photosphere.
In both cases an increased photospheric luminosity is obtained. 
We suggest that the spectral peak in gamma ray bursts is essentially 
due to the Comptonized thermal component from the photosphere, 
where the comoving optical depth in the outflow falls to unity. 
Typical peak photon energies range between those of classical bursts 
and X-ray flashes. The relationship between the observed photon peak 
energy and the luminosity depends on the details of the dissipation, 
but under plausible assumptions can resemble the observed correlations. 

\end{abstract}

\section{Introduction}
\label{sec:intro}

Most GRB models invoke a relativistic outflow, probably channeled into a 
jet, that is energized by a central compact object. The gamma-ray and hard 
X-ray emission is attributed to dissipative processes in the jet. The 
outflow is inferred to be unsteady, on timescales down to a millisecond -- 
indeed internal shocks are the most widely-discussed dissipative process, 
because of their ability to convert bulk kinetic energy into relativistic 
electrons which then radiate (e.g. via synchrotron emission) on very short 
timescales . The outflow would carry baryons, and also magnetic fields 
(which may carry as much power in Poynting flux as does the baryon kinetic 
energy). However, there is another inevitable ingredient of the outflow: 
thermal radiation. This radiation would originate near the base of the 
outflow, where densities are high enough to guarantee (at least 
approximate) thermal equilibration. This thermal radiation would be 
advected outward so long as the jet material remained opaque, and would 
emerge highly collimated, from a `photosphere' where the jet became 
optically thin. 

A laminar and steady jet, viewed head-on, would give rise 
to emission with a thermal spectrum peaking in the hard X-ray or gamma-ray 
band. Moreover, the comoving energy density of this black-body radiation 
could be at least comparable with that of the magnetic field. So, if 
dissipation generates relativistic electrons and supra-thermal pairs, their 
energy losses due to Compton scattering of the thermal radiation would be 
competitive with those from synchrotron emission -- perhaps even dominant. 
Consequently, when dissipation occurs (e.g. via internal shocks) the outcome 
may be a `hardened' (grey body) thermal component, along with a power-law 
component extending to higher photon energies. We suggest that the photon 
energy $E_{pk}$ at which GRB spectra reach a peak may be the (probably 
Comptonized) thermal peak. We discuss how, on this hypothesis, $E_{pk}$ 
would depend on the parameters characterizing the GRB. 

A key parameter in the outflow is plainly the photospheric radius -- 
the radius at which the comoving optical depth along the jet falls to 
unity. In calculating this radius, we must allow for the possibility 
that the electrons associated with the baryons are outnumbered by 
electron-positron pairs (e.g. Eichler \& Levinson, 2000). Moreover,
the number of pairs may be greatly increased by dissipative processes . 
The details depend on whether the photosphere  lies inside 
or outside the saturation radius at which the bulk 
Lorentz factor $\Gamma$ asymptotes to the dimensionless entropy $\eta 
=L_0/({\dot M} c^2)$, where $L_0$ and ${\dot M}$ are the total energy and 
mass outflow rates. For a spherical flow where the free energy emanates 
from a central region $r_0 \sim \alpha r_g =\alpha 2GM/c^2$ comparable to 
the Schwarzschild radius $r_g$ of a central object of mass $M$ (where 
$\alpha\geq 1$), the bulk Lorentz factor grows as $\Gamma(r)\sim r/r_0$ 
outside of $r_0$ up to a saturation radius $r_s\sim r_0\eta$, where it 
saturates at a value $\Gamma\sim\eta$. This simple behavior applies 
for a spherical outflow (or a conical one with jet opening half-angle 
$\theta_j <\Gamma^{-1}$) where there are no internal shocks. We focus 
on this as an illustrative case (bearing in mind that the effective value 
of $r_0$ may be increased by dissipation in the inner jet). Moreover, 
extensions to the cases of convergent or divergent jets are straightforward. 
Inside the saturation radius, the observer-frame photospheric luminosity 
$L_\gamma$ is approximately the total luminosity of the outflow $L_0$, 
since the increasing Doppler boost just cancels the adiabatic decay of the 
comoving characteristic photon energy. On the other hand, if the 
photosphere of an adiabatic flow occurs outside the saturation radius, 
$r>r_s$, the Lorentz factor no longer grows, and $L_\gamma 
(r)=L_0(r/r_s)^{-2/3} < L_0$, the greater part of the energy being in 
kinetic form, $L_k\sim L_0$ (e.g. \Mesz \& Rees, 2000). If this 
photospheric luminosity were the bulk of the observed radiation, the 
radiative efficiency would be low in the latter case.

However, the above scenario can change substantially due to dissipative 
effects such as magnetic reconnection (e.g. Thompson, 1994, Giannos \& 
Spruit, 2004), neutron decay (e.g. Beloborodov, 2003), or internal shocks. 
If the dissipation occurs below the photosphere, the adiabatic decrease of 
the radiative luminosity beyond the saturation radius can be compensated by 
reconversion of some fraction $\epsd \leq 1$ of the kinetic energy into 
radiation, which would re-energize the photospheric component. Moreover, 
dissipation outside the nominal photosphere may lead to sufficient pair 
formation to create a second photosphere, lying outside the original 
nominal photosphere which would have obtained in the absence of 
dissipation.

Thus, if there were sub-photospheric dissipation , the observable 
photospheric luminosity would be boosted by the energy recovered from the 
kinetic energy, which becomes available for converting into radiation or 
pairs. Moreover the dissipated energy would go mainly into Comptonized 
of the thermal radiation advected out from the central engine. Above $r_s$, 
the photospheric luminosity can be boosted to a value $L_\gamma=\epsd L_0 > 
L_0(r/r_s)^{-2/3}$, depending on the dissipation efficiency. We suggest 
that the peak energy of the photon spectrum of gamma ray bursts should be 
identified with the peak of this Comptonized spectrum.

\section{Photospheres, dissipation and pairs} \label{sec:phot}

In the dissipation regions of the flow, all suprathermal or relativistic 
electrons and pairs will lose energy by Compton scattering of the thermal 
radiation (which will be roughly isotropic in the comoving frame). 
Synchrotron losses might dominate for high $\gamma$ electrons, but for 
those with modest $\gamma$ , synchrotron emission will be inhibited by 
self-absorption; these will lose their energy primarily by Compton 
scattering even if the magnetic energy density exceeds that of the thermal 
radiation. They will cool down and thermalize in a time short compared to 
the dynamic time.

Relativistic electrons moving through black-body radiation Compton-boost
each scattered photon by $\gamma^2$, producing a power law rather than
just boosting each photon by a small amount. However, if the slope of
the injected power law is steeper than $-2$, most of the energy will be
at the low energy end, and all the energy of electrons with, say,
$\gamma \lesssim 3$ would go into what would look like a broad thermal
peak. They would emit no synchrotron radiation (because of
self-absorption) and they would not boost any of the thermal
photons by more than a factor $\epsilon_c \sim 10$.

 If the primary dissipation were mainly by strong shocks most of the energy 
might be channeled initially into very high-gamma electrons, which would 
produce photons with a power law spectrum extending to very high energies: 
production of photons above 1 Mev in the comoving frame would only require 
$\gamma\sim 10 -30$ for Compton scattering, and little more than $10^3$ for 
synchrotron emission. However pair production can change this situation, 
leading again to a situation where energy is ultimately dissipated via 
thermal Comptonization. If the compactness parameter is more than unity 
(which, as shown below, is often the case for the usual parameters 
considered) most of the energy in photons with $>$ MeV energies in the 
comoving frame will be converted into pairs with very modest $\gamma$. 
These pairs will then lose their energy (as described above) by Compton 
cooling, resulting in a quasi-thermal spectrum, whose characteristic peak 
is a factor $\epsilon_c \siml 10$ above the original thermal peak, i.e. in 
the tens to hundreds of keV. These pairs establish effectively a new 
photosphere, outside the one that would have been present in their 
absence, and the dissipation (or shocks) responsible for these pairs will 
effectively be a sub-photospheric dissipation, which has different 
characteristics from the more familiar shocks that occur well outside the 
photosphere (e.g. Ghisellini \& Celotti, 1999; Kobayashi, Ryde \& 
MacFadyen, 2002; Pe'er \& Waxman, 2004).

For a GRB outflow of radiative luminosity $L$ and bulk Lorentz factor 
$\Gamma$ in the observer frame, at a radius $r$ the comoving scattering 
opacity due to $e^\pm$ pairs in the high comoving compactness regime is 
\be 
\tau'_\pm \sim \ell'^{1/2} \sim (L \sigma_T/4\pi m_e c^3 \Gamma^3 r)^{1/2}, 
\label{eq:pairopacity} 
\ee 
where $\ell'$ is the comoving frame compactness 
parameter (e.g. Pe'er \& Waxman, 2004). Here we have approximated $L(>1{\rm 
MeV})\sim \epsd L_0$, where $L_0$ is the total luminosity in the observer 
frame, and we have taken other efficiency factors to be of order unity. The 
functional dependence of equation (\ref{eq:pairopacity}) can be obtained 
by considering in the comoving frame (primed quantities, as opposed to 
unprimed quantities in the observer frame) the balance between the rate at 
which pairs annihilate and the rate at which pairs are formed. The latter is 
the rate at which photons capable of pair-producing are introduced into the 
flow, i.e. the photon density above $m_e c^2$ divided by the comoving dynamic 
time, ${n'}_{\pm}^2 \sigma_T c \sim (L/4\pi r^2 m_e c^3 
\Gamma^2)(c\Gamma/r)$, from which follows the pair optical depth $\tau'_\pm 
\sim n'_\pm \sigma_T (r/\Gamma )$. The pair photosphere $r_{ph,\pm}$ is the 
radius where $\tau'_\pm \sim 1$, or 
\be 
r_{ph,\pm} \sim (\epsd /2 \alpha) (m_p/m_e)(L_0 /L_E)\Gamma^{-3} r_0
   \sim 2 \times 10^{14} L_{51}\epsilon_{d,-1}\alpha^{-1}\Gamma_2^{-3}~\cm
\label{eq:pairphot}
\ee
where $L_E=4\pi GM m_p c/\sigma_T\simeq 1.25\times 10^{39} m_1 \erg\si$
is the Eddington luminosity, $r_0=\alpha r_g$ where $\alpha \geq 1$ and
$r_g=2GM/c^2 \simeq 3\times 10^6 m_1~\cm$ is the Schwarzschild radius for
a central object (e.g. black hole) of mass $M\sim 10 m_1$ solar masses,
and $\eta=L/({\dot M}c^2$ is the dimensionless entropy of the relativistic
outflow.

On the other hand, the scattering opacity due to the ordinary electrons 
associated with baryons in the flow would give rise to a 'baryonic 
photosphere' at \be r_{ph,b}\sim (1/2\alpha)(L_0/L_E) \eta^{-1} \Gamma^{-2} 
r_0
        \sim 1.2 \times 
10^{12}L_{51}\alpha^{-1}\eta_2^{-1}\Gamma_2^{-2}~\cm~, 
\label{eq:baryonphot} \ee with the same notation as above.

For an outflow which starts at $r_0 = \alpha r_g =
3\times 10^6 \alpha m_1$ cm, where $\alpha \geq 1$ and the initial
Lorentz factor $\Gamma_0 \sim 1$, under adiabatic conditions
energy-momentum conservation leads  to a Lorentz factor which grows
linearly as $\Gamma(r) \propto r/r_0$ until it reaches a saturation
radius $r_s \simeq r_0\eta \simeq 3\times 10^8 \alpha m_1 \eta_2~\cm$,
beyond which the Lorentz factor saturates to $\Gamma\simeq \eta =$
constant. One can then define two critical limiting Lorentz factors
\ba
\eta_b= &\bigl( \frac{1}{2\alpha}\frac{L}{L_E}\bigr)^{1/4}=
       7.9\times 10^2(L_{51}/\alpha m_1)^{1/4} \cr
\eta_\pm=& \bigl(\epsd \frac{m_p}{m_e}\bigr)^{1/4} ~\eta_b=
       2.9\times 10^3 (L_{51}\epsilon_{d,-1}/\alpha m_1)^{1/4}
\label{eq:etacrit}
\ea
which characterize the behavior of the baryon and pair photospheres below
and above the saturation radius. The pair photosphere behaves as
\be
r_{ph,\pm}/r_0=\cases{
  \eta_\pm    & for $r<r_s$ \cr
  \eta_\pm (\eta/\eta_\pm)^{-3} & for $r>r_s$}
\label{eq:pairphot_eta}
\ee
and the pair photosphere occurs at $r<r_s$ for
$\eta > \eta_\pm$. The baryon photosphere behaves as
\be
r_{ph,b}/r_0 = \cases{
 \eta_b (\eta /\eta_b)^{-1/3} & for $r<r_s$ \cr
 \eta_b (\eta/\eta_b)^{-3}    & for $r>r_s$ }
\label{eq:barphot_eta}
\ee
and the baryon photosphere occurs at $r< r_s$ for
$\eta > \eta_b$. This is shown schematically in Figure 1,
for values of $\alpha =1,~10^4$, i.e.  initial radii
$r_0=3\times 10^6 \alpha_0 m_1 \cm$ and
$r_0= 3\times 10^{10}\alpha_4 m_1 \cm$.

The pair photosphere (equation [\ref{eq:pairphot}]) will be above
the baryon photosphere provided that
\be
\epsd > (m_e/m_p)
\label{eq:etapairabovebar}
\ee
where $\epsd$ characterizes the dissipation efficiency producing
photons above energies $m_e c^2$ in the comoving frame.

\section{Characteristic Photon Luminosities and Temperatures}
\label{sec:lum}

When the conditions of equation (\ref{eq:etapairabovebar}) are satisfied, 
the real (outermost) photosphere is not the baryon photosphere but the pair 
photosphere. The pair photosphere will have a luminosity 
$L_{\gamma\pm}=\epsd L_0\leq L_0$. At $r<r_s$, the observed radiation is 
insensitive to the actual details of the photosphere: the decrease with $r$ 
in the comoving-frame luminosity is compensated by the observer-frame boost 
given by the increasing Lorentz factor $\Gamma$; moreover, there is less 
scope for dissipation (except in the case when Poynting flux far exceeds 
the radiative flux in the jet).

On the other hand, for a photosphere at $r>r_s$ the luminosity 
decays as $L_{\gamma } = L_0(r/r_s)^{-2/3}$ in the adiabatic regime. 
However, if dissipation  occurs  above $r_s$, this leads to an
effective luminosity
\be L_{\gamma} \sim \epsd L_0~. 
\label{eq:Lpm} \ee 
This luminosity is achieved at the baryon photosphere if dissipation 
occurs below this radius, even in the absence of significant pair formation, 
or above the baryon photosphere if dissipation above the baryon photosphere
leads to a pair photosphere radius $r_\pm$ such that $\epsd (r_\pm /r_s)^{2/3}
\geq 1$ (see Fig. 2). In such cases the effective photosphere luminosity 
exceeds what would have emerged from a non-dissipative outflow by 
$\epsd (r/r_s)^{2/3}$. 

The characteristic initial temperature of the fireball outflow is
\be
T_0=(L_0/4\pi r_0^2 c a)^{1/4}
   = 1.2 L_{51}^{1/4}(\alpha m_1)^{-1/2}~{\rm MeV}~,
\label{eq:T0}
\ee
which for a larger $\alpha=10^4$ (i.e. for a larger initial radius
$r_0=3\times 10^{10}\alpha_4 m_1$ cm) would be
$T_0=12.1~L_{51}^{1/4}(\alpha_4 m_1)^{-1/2}~{\rm keV}$.

For $r<r_s$ the observer-frame effective photospheric temperature (even in 
the presence of dissipation) remains as $T_\gamma =T_0$, being boosted by 
the growing Lorentz factor back to its initial value. For $r>r_s$, 
adiabatic effects (in the absence of dissipation) would cause the 
temperature to fall off as $T_\gamma=T_0(r_s/r)^{-2/3}$. 
Since, however, dissipation leads to a luminosity $\epsd L_0$ 
which can exceed the adiabatic value, this results in a temperature 
$T_{\gamma}$ which drops $\propto r^{-1/2}$.  If dissipation is 
maintained all the way to the (baryonic or pair-dominated) 
photosphere, the temperature is 
\be 
T_{\gamma,d}=\epsilon_c \epsilon_d^{1/4}(r/r_s)^{-1/2} T_0~, 
\label{eq:Tpm} 
\ee 
where a factor $\epsilon_c \simg 1$ accounts for possible departures 
from a black-body. This temperature is larger by a factor 
$\epsilon_c\epsilon_d^{1/4}(r/r_s)^{1/6}$ than the adiabatic 
photosphere temperature $T_\gamma$ (Fig. 2).

Thus, one consequence of dissipation is that, even for $\alpha=1$,
e.g. with $\epsd=10^{-1}$ and $r_d/r_s=10^2$, the  characteristic
temperatures can be  $kT_\gamma \sim 60$ keV, i.e. peak photon energies 
in the X-ray flash range, while for $\alpha=10^4$ this energy can 
easily be as low as a few keV.

If dissipation is important only for some range of radii,
starting at $r_s \epsd^{3/2}$ but ceasing, say, at some radius
$r_c$ below the photosphere $r_{ph}$, then the adiabatic decay
$L_\gamma \propto T_\gamma \propto r^{-2/3}$ resumes above $r_c$ until
$r_{ph}$ so the photospheric luminosity would be
lower than implied by eqs. (\ref{eq:Lpm},\ref{eq:Tpm}).

We should note that, in the present context, $r_0$ is essentially the 
radius beyond which the Lorentz factor starts to grow as $\Gamma \propto 
r/r_0$. Thus any dissipation in the inner ``cauldron" or along the inner 
jet in effect pushes out $r_0$. This could come about because of 
entrainment, or because of the oblique shocks that occur when the jet is 
initially poorly collimated (as exemplified in numerical calculations of 
collapsar models such as Zhang \& Woosley 2004, where in effect $r_0\simg 
10^8$ cm). The initial reference temperature $T_0$ is correspondingly 
lower. (Another effect which could change the reference temperature is if 
the inner jet is Poynting-dominated, so that only a small fraction of the 
flux is in the radiation. In this case, pairs can be even more dominant 
inside $r_s$ . There could be modifications to the outflow dynamics if the 
field were tangled and did not obey the straightforward Bernouilli equation 
for a relativistic gas (cf Heinz and Begelman, 2000)).

Dissipation need not necessarily lead to pair formation.
For example, in a ``slow heating" scenario (such as that of Ghisellini 
and Celotti, 1999), the accelerated particles, and the photons associated 
with them, could all have energies substantially below $\sim 0.5$ MeV. 
Dissipation would then not enhance the photospheric radius, but, even so, 
as indicated above, the characteristic photon energies and photospheric 
luminosity could be substantially boosted over what their adiabatic value 
would have been.

An important feature of the model is that millisecond variations 
-- either at the photosphere or due to internal shocks further out -- may 
still be traced back to irregularities in the jet boundary at $r_0$, since 
the characteristic timescale for a nozzle of opening half-angle $\theta_j$ 
is $t_{var}\sim r_0 \theta_j /c$, rather than $r_0/c$ itself, which can be 
less than a millisecond even if $r_o$ is of order $10^8$ cm . 
If internal shocks are to develop, they must be induced by unsteady 
conditions near the base of the jet (resulting in changes in $\eta$ and 
the saturation Lorentz factor). 
While for the usual minimum variability timescale $t_{var}\sim r_0/c$
shocks would develop above the line marked $r_{sh}$ in Fig. 1, for 
$t_{var}\sim r_0 \theta/c$ the shocks can form at radii $r_{sh,j}$ a factor 
$\theta_j$ smaller then for the spherical case, see Fig. 1. Dissipation at 
such or smaller radii is also possible, e.g., in the case of oblique shocks  
induced by irregularities in the walls of the jet, or during the collimation 
of an initially poorly collimated jet, or in the case of dissipation due 
to magnetic reconnection.

Note also that any variability at $r_0$ would alter the conditions 
at the photosphere (and the value of the photospheric radius). Moreover, 
the photospheric changes can be rapid. Obviously this is true if the 
photosphere lies below the saturation radius; however, this condition is 
not necessary, and provided that the photospheric radius is within $\eta^2 
r_0$, there is no smearing of variability on any timescale down to $r_0/c$. 
We would therefore, generically, expect an internal shock to be slightly 
preceded by a change in the luminosity of the thermal component (and in 
$E_{pk}$). Indeed, one is led to conjecture that rapid variations in the 
photosphere could be at least as important as the associated internal 
shocks in causing rapid variability in GRBs. In contrast to shocks, 
variations in the photospheric emission could as readily account for a 
short dip as for a short peak.  Detailed evidence of spectral softening 
during both the rise and fall of individual sub-pulses (c.f. Ryde 2004), 
could clarify the relative contributions of these effects.

\section{Spectrum Formation}
\label{sec:spec}

When dissipation occurs, one expects the photospheric spectrum to be 
"grey" rather than an accurate blackbody, because there would not 
(except near the base of the jet) be processes capable of producing the 
new photons appropriate to a black body with the enhanced energy density. 
All photons emerging from the photosphere will, however, have undergone 
multiple scatterings. 
In the case of shock dissipation, a power law relativistic electron 
energy distribution can be formed, which would upscatter the thermal 
photons into a  power law photon distribution whose index is similar 
to that of synchrotron radiation, pair formation being possible at 
comoving energies $\simg m_e c^2$. At the pair photosphere the comoving 
inverse Compton cooling time is $t'_{IC}\sim 4\times 10^{-3} L_{51}
\epsilon_{d,-1}^2 \alpha^{-2}\gamma_{e,3}^{-1}$ s, while the dynamic time 
is $t'_{dyn}\sim 3\times 10^1 L_{51}\epsilon_{d,-1}\alpha^{-1} \eta_2^{-4}$ s.
The interplay between the electron and photon distributions requires a 
detailed analysis, and is discussed in Pe'er, et al, 2005b.  
For a slow heating scenario, such as that of Ghisellini and Celotti 
(1999) but with the added feature of dissipation (e.g. from magnetic 
reconnection, or from multiple shocks and/or MHD turbulence behind them), 
one expects the electrons to be heated to more modest values, say 
$\gamma_e \siml$ few, but the electrons could keep being reheated 
every Compton cooling time. In this case pair formation is at best 
modest (Pe'er, et al, 2005b), so the effects outside the baryonic 
photosphere are not significant. If slow dissipation occurs at or 
below the baryonic photosphere, where pair formation is suppressed, 
the IC cooling time is $t'_{IC}\sim 10^{-4} L_{51}\alpha^{-1}
\gamma_{e,0.5}^{-1}\eta_2^{-4}$ s while the dynamic time is 
$t'_{dyn}\sim 2\times 10^{-1} L_{51}\alpha^{-1} \eta_2^{-4}$ s. 
The dissipative baryonic photosphere thermal peak is at $3kT_{\gamma,b}
\sim 20 L_{51}^{-1/2}\epsilon_{d,-1}^{1/4}\alpha^{1/2}\eta_2^2$ keV, 
which (depending on $\gamma_e$) may get upscattered by factors $\sim 1-10$.
The schematic shape of the spectrum is shown in Fig. 3 (c.f. Pe'er et al, 
2005a, 2005b), showing the original quasi-thermal Wien component, 
the up-scattered photospheric component resulting from sub-photospheric 
dissipation and Comptonization, and a possible additional synchrotron 
component from shocks outside the photosphere. The peak frequency scales 
with the amount of dissipation according to a power law which depends
on how many new photons are produced. (The photon production depends on the 
radial dependence of the dissipation and on the detailed dissipation mechanism.)

The dependence of the spectral peak energy on the burst parameters,
as observed in a given energy range by a given instrument,  depends on
the specific mechanism responsible for the spectrum in that energy range.
In the BATSE and Beppo-SAX energy range (roughly 20 keV to 0.5 MeV),
there is a quantitative relationship observed between the spectral
peak energy $E_{pk}$ and the isotropic-equivalent luminosity of the
burst in tat energy range, $L_{iso}$ (which requires a knowledge of
the redshift of the burst). This relationship (Amati et al, 2001) is
$E_{pk}\propto L_{iso}^{1/2}$, for a score of bursts with redshifts.
Our generic assumptions naturally yields a peak in the relevant range, 
but cannot predict any correlation with $L$ without a more specific model
which relates $L$ to the other significant parameters, in particular 
$r_0$ and $\eta$. Without going into details, we may consider several
possibilities.

If one seeks to explain this relationship by interpreting the peak
energy as the synchrotron peak in a simple standard internal shock
scenario, one expects the dependence (e.g.  Zhang \& \Mesz, 2001)
\be
E_{pk} \propto \Gamma^{-2} t_{var}^{-1}L^{1/2} ~,
\label{eq:Epkintsh}
\ee
where $L_{\gamma, iso} \sim L$. Here $\Gamma$ and $t_{var}$ are
the Lorentz factor of the outflow and its typical variability
timescale. If the  latter two quantities are approximately the same
for all bursts, this would reproduce the Amati et al (2001) relation.
However, it is not obvious why there should be a constancy of $\Gamma$
and $t_{var}$ across bursts, even if approximate.

If the spectral peak is of a quasi-thermal origin
determined by the photosphere (possibly shifted up by Comptonization,
e.g. from pair dissipative effects such as discussed above), and if
there are enough photons to guarantee an approximate black body
distribution, the peak photon energy in the observer frame is,
using equation (\ref{eq:pairphot}),
\be
E_{pk}\propto \Gamma kT'_{pk} \propto \Gamma (L/\Gamma r^2)^{1/4}
       \propto \Gamma^2 L^{-1/4} \propto L^{(8\beta -1)/4},
\label{eq:EpkBB}
\ee
which depends mainly on the Lorentz factor $\Gamma$. If the latter
in turn depends on $L$, e.g. as $\Gamma\propto L^{\beta }$, one obtains
the last part of equation (\ref{eq:EpkBB}).
For instance, taking the observed Frail et al (2001) relation
$L_{\gamma, iso}\propto \theta^{-2}$ between the jet opening half-angle
$\theta$ inferred from the light-curve break, and using the causality
relation $\theta \sim \Gamma^{-1}$, equation (\ref{eq:EpkBB}) becomes
$E_{pk}\propto L^{3/4}$.

If dissipation occurs mainly very close to the central engine, this
could result in a larger radius $r_0$, where $r_0$ is defined as the
radius beyond which $\Gamma\propto r/r_0$. Assuming that the ``drag"
or dissipation at the base increases $r_0$ according to, e.g.,
$r_0\propto L^{-\beta'}$, for a photosphere occurring inside the
saturation radius, $r_0 < r_{ph,\pm} <r_s$, the growth of the
Lorentz factor $\Gamma \propto r/r_0$ cancels out the adiabatic
drop $T'\propto r^{-1}$ of the comoving temperature, and one has
\be
E_{pk}\propto r_0^{-1/2} L^{1/4} \propto L^{((2\beta' +1)/4}~.
\label{eq:Epksat}
\ee
Hence, for $\beta'=(0.5,~1)$ one has $E_{pk}\propto (L^{1/2},~L^{3/4})$.

In the extreme 'photon starved' case (likely to apply if the dissipation 
is concentrated not far inside the photosphere) where the photon number 
$N_\gamma$ is constant), one would have $E_{pk} \propto L/ N_\gamma \propto 
L$. 

Thus, a variety of $E_{pk}$ vs. $L$ dependences might in principle be 
expected, depending on the uncertain physical conditions just below the 
photosphere, some of which approximate the reported $L^{1/2}$ behavior. 
Ghirlanda et al (2004) have recently claimed an empirical correlation 
between $E_{pk}$ and a different quantity, the angle-corrected total energy 
$E_{tot}= E_{iso}(1-\cos\theta_j)\sim E_{iso}(\theta_j^2/2)$, where 
$E_{iso} \simeq L_{iso} t_{\gamma}$ and $t_\gamma$ is the burst duration. 
They find a tighter correlation between $E_{pk}$ and $E_{tot}$ than between 
$E_{pk}$ and $E_{iso}$, for bursts with observed redshifts and breaks. 
Furthermore, in contrast to the Amati et al (2001) $E_{pk}\propto 
E_{iso}^{1/2}$ dependence, they deduce from the data a steeper slope, 
$E_{pk}\propto E_{tot}^{0.7}$. Taking a standard burst duration and jet 
opening angle, this is of the form discussed in equation (\ref{eq:EpkBB}) 
or (\ref{eq:Epksat}). A critique of the methods for obtaining both types of 
correlations from the data is given by Friedman and Bloom (2004). We should 
note that such correlations are generally derived assuming that the 
efficiency of gamma-ray production is the same for all bursts, 
independently of the luminosity or the total energy. If, however, the 
efficiency were lower for the weaker (and therefore softer) bursts, then 
the correlation would have a flatter $E_{pk}$ vs. $E_{tot}$ slope than 
currently derived from the data. This is because, for a given gamma-ray 
isotropic luminosity, the momentum outflow per unit solid angle would be 
higher than they assume. This means that the standard jet-break argument 
would imply a narrower beam than inferred under the constant efficiency 
assumption, and therefore a lower $L_{tot}$ (for a given $E_{pk}$) than the 
values currently derived.

\section{Discussion}
\label{sec:disc}

We have considered dissipative effects below the photosphere of GRB
or XRF outflows, such as, e.g., due to magnetic reconnection or shocks.
Such dissipation can lead to copious pair formation, dominating the
photospheric opacity.  Alternatively, if dissipation occurs not too far
above an initial photosphere, it can result in a second effective
photosphere, situated outside the initial one.

Sub-photospheric dissipation can increase the radiative efficiency of the 
outflow, significantly boosting the quasi-thermal photospheric component, 
so that it may well dominate the much-discussed synchrotron component from 
nonthermal shocks outside the photosphere. The hypothesis that GRB emission 
is dominated by a Comptonized thermal component offers a natural 
explanation for the thermal GRB spectra discussed most recently, e.g., by 
Ryde (2004). It can also naturally explain the steeper than synchrotron 
lower energy spectral indices (Preece, et al, 2000; Lloyd, Petrosian \& 
Mallozi, 2000) noticed in some bursts.

The quasi-thermal peak of the photospheric spectrum is controlled by
the total luminosity $L_0$ and by the reference injection radius $r_0$
above which the Lorentz factor starts to grow linearly. Dissipation
near the central object or along the inner jet can result in an
increase $r_0$, thus lowering the reference temperature of the outflow
which characterizes the quasi-thermal photospheric component. The
characteristic variability timescales $r_0\sin\theta/c$ for jets
with the observationally inferred opening half-angles $\theta$ are
in the millisecond range. The spectral peak of the dissipation-enhanced
photospheric component, upscattered in energy by factors of $\sim 10$
due to electrons accelerated in the dissipation process, results in typical
photon energies ranging between those of classical bursts and X-ray flashes.

The relationship between the observed photon peak energy and the luminosity 
can have a variety of functional forms, which depends on a number of so-far 
poorly determined parameters. However, plausible assumptions can lead to 
relationships of the type $E_{pk}\propto L_{iso}^{1/2}$ (Amati, et al, 
2000), or $E_{pk}\propto E_{tot}^{0.7}$ (Ghirlanda et al, 2004). Even 
though more physics and a more specific model will be needed before we can 
explain the correlations, the idea that $E_{pk}$ is essentially a thermal 
peak seems more readily able to account for a `standardized' value in a 
given class of objects, because there is not a steep $\Gamma$-dependence 
(and indeed to first order the $\Gamma$ factor cancels out, because 
adiabatic cooling in the comoving frame is compensated by the Doppler 
boosting).

In summary, our main result is that  a spectral  peak at photon  energies 
in the range of tens to hundreds of keV, typical of XRFs and GRBs, can 
naturally arise from an outflowing jet, in which dissipation below a 
baryonic or pair-dominated photosphere enhances the radiative efficiency 
and gives rise to a Comptonized thermal spectrum. On this hypothesis, 
the recently-discovered correlations between $L$ and $E_{pk}$ would be 
an important diagnostic of how the key jet parameters --  physics near 
the 'sonic point', baryon contamination, etc -- depend on $L$.

\acknowledgements
Research supported in part by NASA NAG5-13286, the Royal Society and the
Monell Foundation. We are grateful to S. Kobayashi, D. Lazzati, A. Pe'er
and the referee for useful comments.

\clearpage
\begin{figure}[htb]
\centering
\epsfig{figure=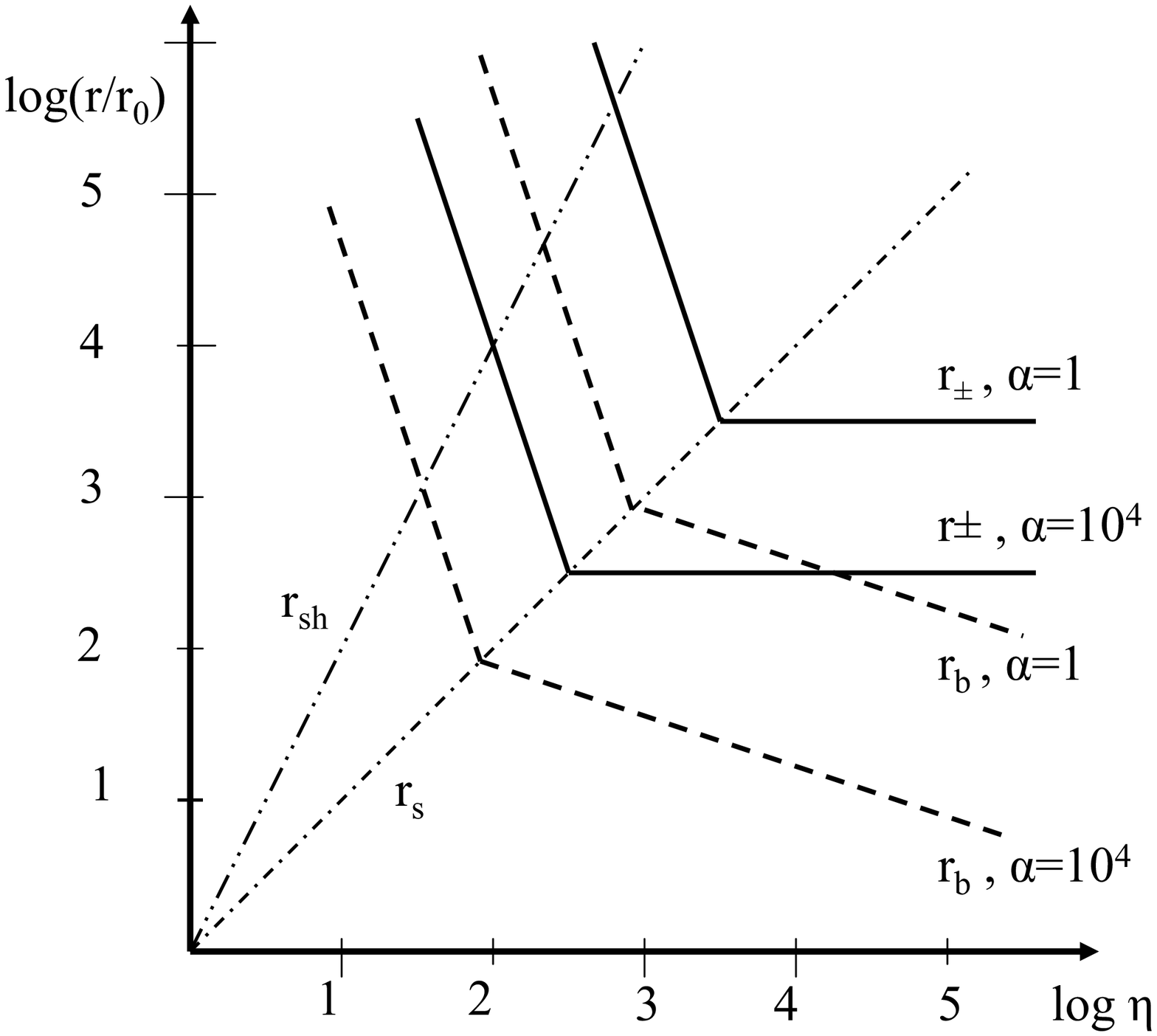,width=5.in, height=4.in}
\figcaption{
Radii of the pair (full) and baryon (dashed) photospheres as a
function of $\eta$ for $\epsd=10^{-1}$, $L_0=10^{51}$ erg/s,
$\alpha=1$ ($r_0=3\times 10^6 m_1\cm$) and $\alpha=10^4$ ($r_0=
3\times 10^{10}m_1\cm$).  Also shown are the saturation radius
$r_s=r_0\eta$, and the spherical minimum shock radius $r_{sh}=
r_0\eta^2$. Instabilities at the nozzle $\theta$ of a jet could
lead to shocks at a lower mininum radius $r_{sh,j}$, while magnetic 
dissipation could in principle occur both above and below $r_s$.
}
\label{fig:photeta}
\end{figure}
\begin{figure}[htb]
\centering
\epsfig{figure=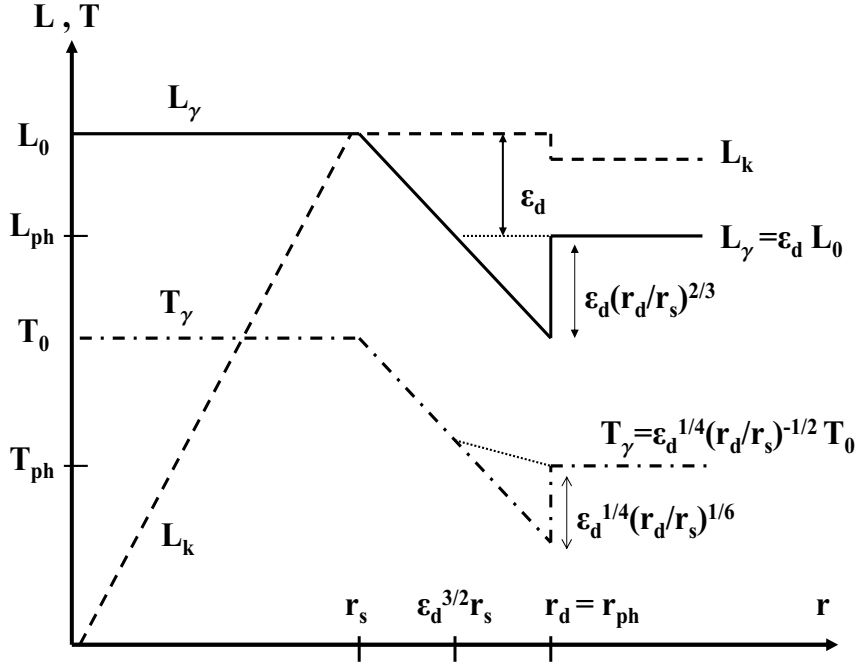,width=5.in,height=4.in}
\figcaption{
Kinetic luminosity and photospheric radiation luminosity as a function
of radius. Beyond the saturation radius the luminosity decays as
$L_\gamma\propto r^{-2/3}$, but beyond $\epsilon_d^{3/2}r_s$ the
fraction $\epsilon_d$ of the kinetic energy reconverted into
radiative (pair) form becomes significant. Also shown is the value
of the observer temperature. Comptonization at the pair photosphere
(see text) could boost this by an additional factor $\siml 10$.
}
\label{fig:lumrad}
\end{figure}
\begin{figure}[htb]
\centering
\epsfig{figure=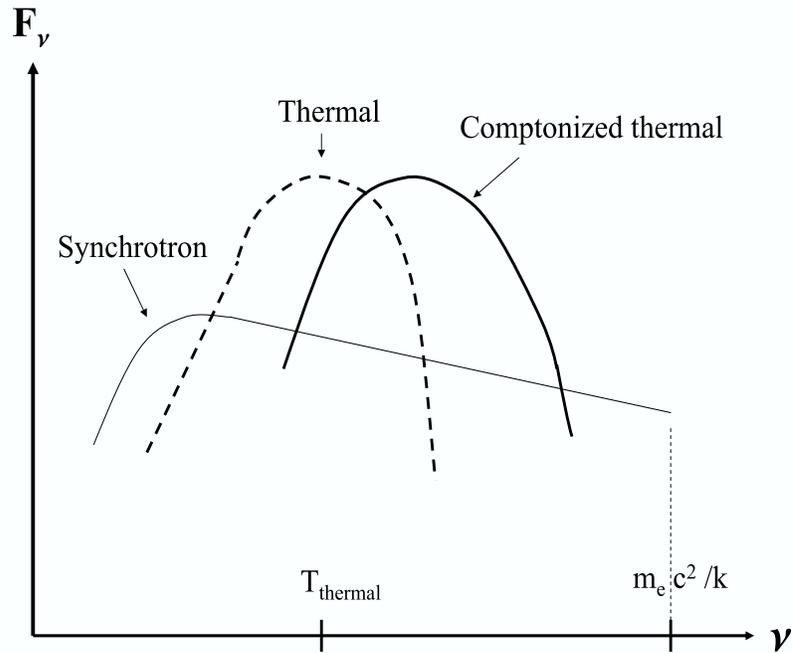,width=5.in,height=4.in}
\figcaption{Schematic comoving frame spectrum, showing the photospheric
(thermal) spectrum and its Comptonized  component, as well as a shock
synchrotron component (assumed to arise further out). This is the
generic spectrum characterizing a slow dissipation model (see text).
Shocks with pair formation could lead to an additional component
at higher energies.
}
\label{fig:spectrum}
\end{figure}

\end{document}